\documentclass{aa}
\usepackage{amsmath}
\usepackage{amssymb}
\usepackage[T1]{fontenc}
\usepackage{graphicx}
\usepackage{natbib}
\usepackage{type1cm}

\bibpunct{(}{)}{;}{a}{}{,}

\newcommand{\?}{\mbox{(?)}\marginpar{\center\hspace{0pt}?}\typeout{WARNING: question}}

\begin{document}

   \title{Three-dimensional simulations of type Ia supernovae}

   \author{M.\ Reinecke \and
           W.\ Hillebrandt \and
           J.C.\ Niemeyer
          }

   \offprints{M.\ Reinecke \\({\itshape martin@mpa-garching.mpg.de})}

   \institute{Max-Planck-Institut f\"ur Astrophysik,
              Karl-Schwarzschild-Str.\ 1, 85741 Garching, Germany}

   \date{Received Apr 25, 2002; accepted Jun 11, 2002}

   \abstract{
     We present the results of three-dimensional
     hydrodynamical simulations of the subsonic thermonuclear burning phase in type Ia
     supernovae. The burning front model contains no adjustable
     parameters so that variations of the explosion outcome can be
     linked directly to changes in the initial conditions. In
     particular, we investigate the influence of the initial flame
     geometry on the explosion energy and find that it appears to be weaker
     than in 2D. Most importantly, our models predict global
     properties such as the produced nickel masses and ejecta
     velocities within their observed ranges without any fine tuning.

   }
   \maketitle

\section{Introduction}

In a series of papers \citep{reinecke-etal-99a, reinecke-etal-99b, 
reinecke-etal-02a} we described a new numerical tool to simulate
thermonuclear explosions of white dwarfs in two and three spatial
dimensions. Our aim was to construct models of type Ia supernova (SN Ia)
explosions that are as free from non-physical parameters as currently
feasible. Thermonuclear burning, in particular, is represented by a
subsonic turbulent flame whose local velocity is derived from a subgrid-scale model
for unresolved turbulent fluctuations. Solved in combination with the
compressible Euler equations, this model contains no free parameters that
could be adjusted in order to fit SN Ia observations. Consequently,
the initial white dwarf model (composition, central density, and
velocity structure), as well as assumptions about the location, size and shape
of the flame surface as it first forms fully determine the simulation
results. Here, we concentrate on variations of the latter.

In \cite{reinecke-etal-99a}, we confirmed the
earlier result of \cite{niemeyer-etal-96} that, at least in 2D, the 
explosion energy and amount of $^{56}$Ni produced in the explosion
(which determines the brightness of an SN Ia) are sensitive 
to the ignition conditions. To be more precise, a more complicated
topology of the initial nuclear flame seems to lead to higher 
Ni-production and, consequently, more powerful explosions. One might
even speculate that the randomness of the ignition process could be 
the reason for the observed spread in properties of normal SNe Ia
\citep{hillebrandt-niemeyer-2000}. 
  
This article continues the presentation of numerical simulations of SNe
Ia by testing the effect of
different initial conditions on the simulation outcome in three
dimensions. In this context, the simultaneous runaway at 
several different spots in the central region of the progenitor star 
is of particular interest. A plausible ignition scenario was suggested 
by \cite{garcia-woosley-95}.

All simulations were performed using the algorithms presented by
\cite{reinecke-etal-02a, reinecke-etal-99a}.
The initial model for the white dwarf,
assumed to be at the Chandrasekhar mass, as well as the grid 
geometry and symmetry assumptions, are identical to the setup 
described in Sects.\ 3 and 4 of \cite{reinecke-etal-02a}.

Sect.\ \ref{multipoint} of this paper presents the initial front geometry
and its temporal evolution for two multi-point ignition calculations.
Various aspects of all three-dimensional calculations are then compared
in Sect.\ \ref{seccomp3d}. As far as possible, this comparison is then
extended to other existing results of SN Ia simulations. In the
case of parameterized one-dimensional calculations such comparisons are
difficult, since most of the published data have no analogy in three
dimensions. As an alternative, we suggest to track and compare the
amount of burned material as function of the density at which the
burning took place. 

Generally speaking, we confirm the earlier 2D results: models with
more, or better resolved, ignition spots tend  
to produce more radioactive Ni also in 3D, although the effect is
somewhat smaller. But together with the gain of Ni by going 
from 2D to 3D our models predict amounts that are in good
agreement with those inferred from the observations
\citep{contardo-etal-00}. This will also be discussed in Sect.\
\ref{seccomp3d}, and our conclusions follow in Sect.\ \ref{discuss}.

\section {Multi-point ignition scenarios}
\label{multipoint}

  \begin{figure}[tbp]
    \centerline{\includegraphics[width=0.48\textwidth]{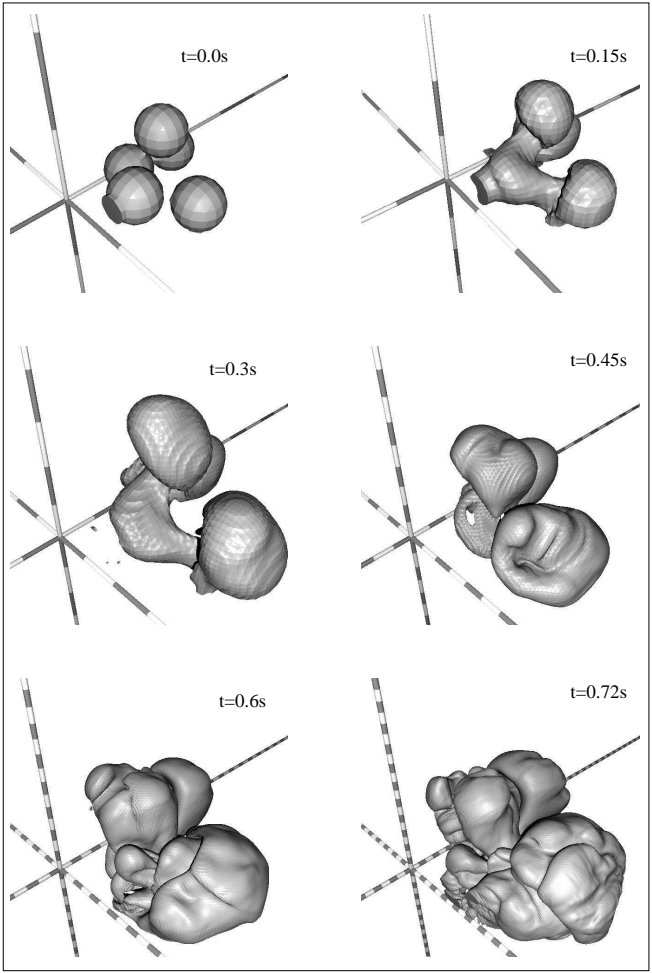}}
    \caption{Snapshots of the flame front for scenario b5\_3d. The fast merging
       between the leading and trailing bubbles and the rising of the entire
       burning region is clearly visible. One ring on the coordinate axes
       corresponds to $10^7$cm.}
    \label{b5_3d_front}
  \end{figure}
  \begin{figure}[tbp]
    \centerline{\includegraphics[width=0.48\textwidth]{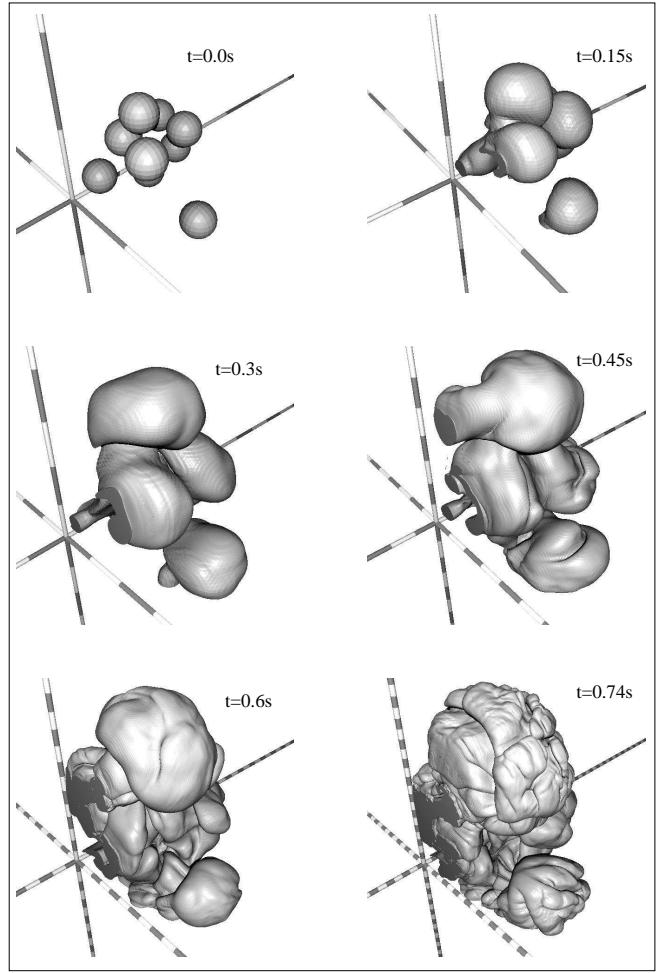}}
    \caption{Snapshots of the flame front for the highly resolved scenario
    b9\_3d. One ring on the coordinate axes corresponds to $10^7$cm.}
    \label{b9_3d_front}
  \end{figure}

Two different calculations were performed to investigate the off-center ignition
model in three dimensions. The simulation b5\_3d\_256 was carried out on a grid
of 256$^3$ cells with a central resolution of 10$^6$\,cm and contained five
bubbles with a radius of $3\cdot 10^6$\,cm, which were distributed randomly
in the simulated octant within $1.6\cdot10^7$\,cm of the star's center.
As an additional constraint, the bubbles were required not to overlap
significantly with
the other bubbles or their mirror images in the other octants. This maximizes
the initial flame surface.
The exact algorithm for the positioning of the bubbles was the following:
First, the centers of the bubbles were chosen randomly within the radius
of $1.6\cdot10^7$\,cm. A realization was accepted when the distance between
any two bubble centers (including the mirror images across the coordinate
planes) was larger than 1.3 bubble radii.
A particular realization of these initial
conditions and its temporal evolution is shown in Fig.\ \ref{b5_3d_front}.
The locations of the bubble centers are listed in Table \ref{b5_table}.

\begin{table}
\begin{center}
\begin{tabular}{|c|r|r|r|r|r|}
\hline
Part.\ No.& 1&2&3&4&5 \\ \hline
x-Pos. & 128 & 35 & 42 & 85 & 63 \\ \hline
y-Pos. & 76 & 60 & 129 & 27 & 85 \\ \hline
z-Pos. & 27 & 30 & 30 & 41 & 83 \\ \hline
\end{tabular}
\end{center}
\caption{Approximate position of the bubble centers in the model b5\_3d\_256.
All lengths are given in kilometers.}
\label{b5_table}
\end{table}

In a way very similar to the evolution of earlier two-dimensional simulations
(cf.\ Fig.\ 6 of \citealt{reinecke-etal-99b}) the flame kernels closer to the
center are
elongated very rapidly and connect to the outermost bubbles within 0.15\,s.
The whole burning region disconnects from the coordinate planes and starts
to float slowly towards the stellar surface. 

In an attempt to reduce the initially burned mass as much as possible without
sacrificing too much flame surface, the very highly resolved model b9\_3d\_512
was constructed. It contains nine randomly distributed, non-overlapping
bubbles with a radius of 2$\cdot$10$^6$\,cm within
$1.6\cdot10^7$\,cm of the white
dwarf's center. To properly represent these very small bubbles, the cell
size was reduced to $\Delta=5\cdot 10^5$\,cm, so that a total grid size of
$512^3$ cells was required.

Starting out with very small flame kernels is desirable, because
the initial hydrostatic equilibrium is better preserved if only a little
mass is burned instantaneously. Furthermore the floating
bubbles are expected to be even smaller in reality than in the presented model
($r\lessapprox5\cdot10^5$\,cm, see \citealt{garcia-woosley-95}).

\begin{table}
\begin{center}
\begin{tabular}{|c|r|r|r|r|r|}
\hline
Part.\ No.& 1&2&3&4&5 \\ \hline
x-Pos. & 70.2 & 106.4 & 22.8 & 77.8 & 137.2 \\ \hline
y-Pos. & 48.6 & 24.6 & 121.2 & 106.7 & 68.2 \\ \hline
z-Pos. & 128.0 & 105.2 & 77.4 & 88.3 & 20.4 \\ \hline\hline
Part.\ No.& 6&7&8&9& \\ \hline
x-Pos. & 47.7 & 41.4 & 27.5 & 42.2 & \\ \hline
y-Pos. & 45.5 & 133.1 & 24.1 & 85.0 & \\ \hline
z-Pos. & 91.2 & 34.0 & 39.3 & 21.7 & \\ \hline
\end{tabular}
\end{center}
\caption{Position of the bubble centers in the model b9\_3d\_512. All
lengths are given in kilometers.}
\label{b9_table}
\end{table}

The center locations of our particular realization are given in
Table \ref{b9_table}.
Snapshots of the front evolution (Fig.\ \ref{b9_3d_front}) exhibit features
very similar to those observed in the lower-resolution 3D models. Only in the
last plot the formation of additional small-scale structures becomes evident.

  \begin{figure*}[tbp]
  \centerline{\includegraphics[width=0.8\textwidth]{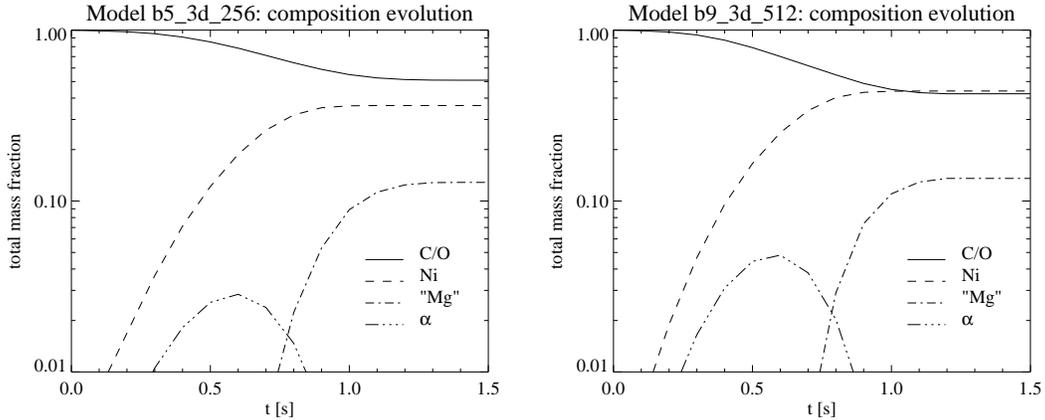}}
  \caption{Time evolution of the chemical composition in the models
  b5\_3d\_256 and b9\_3d\_512.}
  \label{nuc_evolve3d}
  \end{figure*}

The direct comparison of the nuclear evolution with model b5\_3d\_256
in Fig.\ \ref{nuc_evolve3d} nevertheless reveals differences in the early
explosion stages:
the high-resolution simulation produces slightly more nickel and considerably
more $\alpha$-particles during the first half second. This phenomenon is
most likely explained by the discrepancy in the ratio between the
initial flame surface and the volume of burned material in the two models.
The higher burned mass in the five-bubble model initiates an early bulk
expansion of the star and therefore causes a rapid drop of the central density.
Since its flame surface is quite small compared to the burned volume,
only relatively little mass can be burned at high densities.
In the nine-bubble model, on the other hand, the star expands more slowly at
first because less material is burned instantaneously,
and the mass fraction of $\alpha$-particles in the reaction products
is consequently rather high. Since the energy buffered in those
$\alpha$-particles is not immediately used to drive the expansion,
the flame can consume more material at higher densities and has more time
to increase its surface as a result of hydrodynamical instabilities.

\section{Comparison of all 3D simulations}
\label{seccomp3d}
\subsection{Energy release}
  \begin{figure}[tbp]
    \centerline{\includegraphics[width=0.48\textwidth]{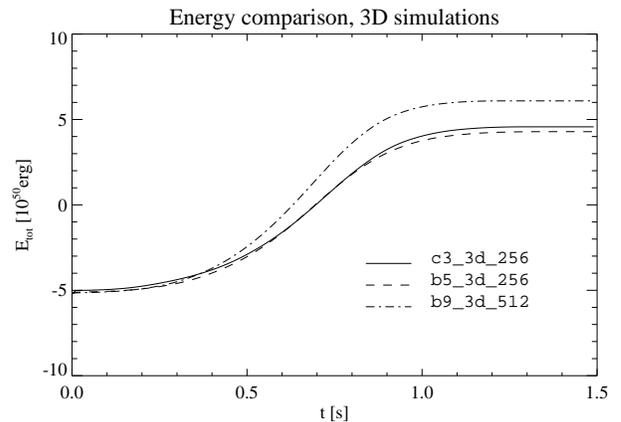}}
    \caption{Energy evolution of the three-dimensional explosion
    models of Fig. 1 (dashed) and Fig. 2 (dashed-dotted). For
    comparison we also show the centrally ignited (``three fingers'') model of
    \cite{reinecke-etal-02a} (solid line).
    While the centrally ignited and the five-bubble models are remarkably
    similar, the high-resolution nine-bubble simulation reaches higher energies
    despite relatively slow initial burning.}
    \label{comp3d}
  \end{figure}

In addition to the two simulations presented above, the following
comparison also includes the model c3\_3d\_256, which was described
in \cite{reinecke-etal-02a}. It must be emphasized that the comparison
between the normal and high-resolution calculations can be qualitative
at best, because the initial conditions as well as the resolution of
model b9\_3d\_256 are different from the other two, and it was not possible
to perform a 3D resolution study like the two-dimensional one discussed
in \cite{reinecke-etal-02a} in order to investigate the influence of
the cell size on the simulation outcome. However, if one assumes that
numerical convergence is reached as long as the flame stays in the uniform
grid region and the cell size is at most 10$^6$cm (which seems very plausible
according to the 2D study), at least the evolution during the initial
$\approx$0.8~seconds should be practically independent of the employed
resolution. For example, if the model b5\_3d would be re-calculated at
increased resolution, the authors would expect identical behaviour to the
original simulation up to 0.8~seconds after ignition, and probably
slightly increased energy generation afterwards.

During the first 0.5 seconds, the three models
are nearly indistinguishable as far as the total energy is
concerned (see Fig.\ \ref{comp3d}), which at first glance appears
somewhat surprising, given the quite different initial conditions.
A closer look at the energy generation rate actually reveals noticeable
differences in the intensity of thermonuclear burning for the three
simulations, but since the total flame surface is initially very small,
these differences have no visible impact on the integrated curve in the
early stages.

However, after about 0.5 seconds, when fast energy generation sets in, 
the nine-bubble model burns more vigorously due to its larger surface 
and therefore reaches a higher final energy level. Fig.\ \ref{comp3d}
also  shows that the centrally ignited model c3\_3d\_256 
of \cite{reinecke-etal-02a} is almost identical to the off-center
model b5\_3d\_256 with regard to the explosion energetics.

Obviously, the scatter in the final energies due to different initial 
conditions appears to be substantially smaller than for 
simulations in two dimensions (see \citealt{reinecke-etal-99a}).
This is, in part, good news, since it demonstrates the robustness of
the explosion mechanism. On the other hand, it may lower the chances
to explain all ``normal'' type Ia supernovae by random variations
of ignition conditions alone. Whether this result is a consequence 
of the particular choice of initial conditions or the self-regulation 
mechanism described in \cite{reinecke-etal-02a} works more efficiently 
in 3D needs to be studied further.

\subsection{Distribution of the ejecta in real and velocity space}
\label{asymmetry}

  It is deduced from observations that the angular distribution of the
  ejecta is fairly homogeneous in SN Ia. This applies to the material
  velocities as well as the chemical composition. Concerning the
  ejecta speeds, the performed simulations are in good agreement
  with observations, since the subsonic nature of the combustion process
  allows for fast equilibration of pressure fluctuations and therefore
  generates a practically spherical expansion. However, the final 
  distribution of the reaction products consists of a few 
  large ``lumps'' of processed material separated 
  by areas of unburned carbon and oxygen. Similar results were 
  also obtained by \cite{niemeyer-hillebrandt-95a}.
  This apparent inconsistency with observations is possibly explained
  by the still limited numerical resolution, which allows only a few initial
  features like bumps in the front or burning bubbles to be prescribed.
  These in turn will dominate the large-scale front geometry at late times.
  In reality the number of bubbles is likely to be much larger, so that the
  initial conditions could be isotropic in a statistical sense, which would
  result in a much more uniform ejecta composition.
  The discussed constraints for the initial conditions in the models b5\_3d
  and b9\_3d also caused the coordinate planes to be practically free of burned
  material. As a consequence, the most likely macroscopic flow pattern that
  will develop consists of an updraft near the center\? of the octant and
  downflow at the edges of the computational domain.

  \begin{figure*}[tbp]
  \centerline{\includegraphics[width=0.8\textwidth]{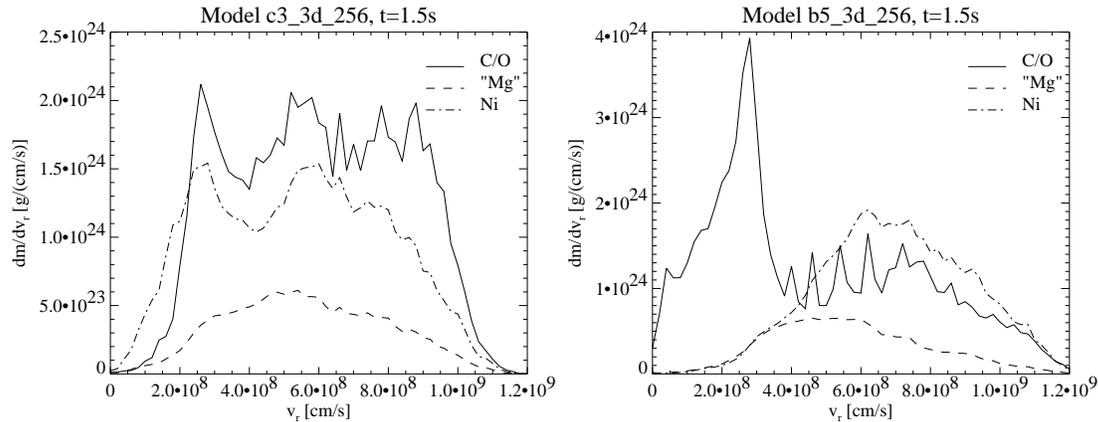}}
  \caption{Distribution of chemical species in radial velocity space
    for models c3\_3d\_256 and b5\_3d\_256 after 1.5 seconds.
    While the elements are more or less
    uniformly distributed in the centrally ignited model, the floating-bubble
    model still contains large amounts of unprocessed material near the center.}
  \label{velprofile}
  \end{figure*}

  The time evolution of line intensities in SN Ia spectra and also
  the line widths due to the Doppler broadening can be used to determine the
  radial position of different elements in the supernova and their
  velocity. A rough estimate for the velocity distribution of the ejecta
  for the models c3\_3d\_256 and b5\_3d\_256 is shown in
  Fig.\ \ref{velprofile}. It must be stressed that these graphs 
  do not represent the final ejecta speeds, since 1.5 seconds after 
  ignition --- at the end of the simulation --- the
  gravitational binding energy is still rather high, and leaving the
  potential well will cause some slowdown, especially in the inner regions.
  Nevertheless the maximum speed and the relative distribution of the elements
  in velocity space allow for some conclusions concerning the 
  predictive power of our simulations.

  The ejected material reaches speeds of up to 12\,000\,km/s for both initial
  conditions which is well within the observed range of 10\,000 to 
  15\,000\,km/s of typical SNe Ia. However, the
  composition of the ejecta in velocity space is quite different from
  results obtained in one-dimensional simulations, insofar as both fuel and ashes
  are present at almost all velocities. Such a situation is, by definition,
  impossible in centrally ignited one-dimensional model, because
  everything inside the radius of the flame must have been burned and the
  material further out cannot have been processed. Multidimensional
  calculations, in contrast, allow simultaneous burning at many different
  radii, and fuel and ashes can be stirred by large-scale 
  turnover motions.

  Whether the velocity profiles of the current simulations are compatible
  with observations has not been thoroughly investigated so far. The large
  amount of carbon and oxygen at low velocities which is typical for the
  floating-bubble calculations (e.g.\ in the right hand plot of
  Fig.\ \ref{velprofile}) has not been detected yet in supernova 
  spectra and, therefore, might not be present in real SN Ia.
  Whether this feature is typical for offcenter ignition scenarios or caused
  by our prescriptions for the initial flame geometry
  (see also Sect.\ \ref{asymmetry}) must be investigated in further
  calculations.
  On the other hand, it is currently conjectured that a certain amount 
  of unburned material could remain undetected 
  by current spectroscopy near the center of the remnant (P.\ Mazzali,
  personal communication), so that the off-center ignition scenario
  cannot be firmly ruled out.

\subsection{The final chemical composition}

Our simulations only employed a minimal nuclear reaction network which
was sufficient for a good approximation of the thermonuclear energy release.
For that reason, the produced abundances cannot be taken literally, but must
be interpreted carefully when trying to derive light curve shapes or spectra
from our results.

In our simulations, the abundance of $^{56}$Ni actually represents the
sum of all Fe-group elements synthesized during the explosion.
Due to the very similar nuclear binding energies of all those nuclei,
this simplification is justified as far as the explosion energy is concerned.
To calculate the light curve, however, $^{56}$Ni and the rest of the
Fe-group elements play significantly different roles. While the radioactive
decay of $^{56}$Ni supplies the energy for the light emitted by the supernova
ejecta and thereby determine the bolometric luminosity, the presence of the
other nuclei enhances the opacity of the ejecta and causes a broadening of
the light curve.

It is expected that ``real'' $^{56}$Ni accounts for $\approx70-80$\% of the
Fe-group elements \citep{thielemann-etal-86}, 
which means that the corrected $^{56}$Ni mass
is still in rather good agreement with the amounts postulated by
\cite{contardo-etal-00}. However, since the relative mass fraction of
$^{56}$Ni and the total Fe-group elements may also vary in space, reliable
answers can only be given once detailed post-processing using a large
reaction network has been performed on the existing data. This
work is currently in progress.

\section {Comparison to previous simulations}

It has been shown above that, as far as comparisons are presently possible,
our numerical models produce results which agree fairly well with real 
SN Ia explosions. Unfortunately, the only parameters we can directly
and quantitatively compare to the observations at present  
are the energy release and -- to some extent -- the nickel masses;
all other data derived from the simulations (like the scatter in the final
energies and the amount of intermediate-mass elements) are of qualitative
nature only. Although they agree roughly with expectations, they may
not provide a criterion for the validity of the model. In order to
allow for a more detailed comparison with the observations the code
has to be extended. Apart from post-processing the ejecta to produce
detailed nuclear abundances, it might also be necessary to model
the complex physical processes taking place in the later SN stages,
up to several weeks after explosion.

As an intermediate step, it might therefore be useful to perform a detailed
comparison of our results with one of the successful phenomenological SN Ia
models like Nomoto's W7 \citep{nomoto-etal-84}. These one-dimensional 
models contain complex nuclear reaction networks and have been used 
to generate synthetic spectra and light curves.
A good candidate for such a comparative study would be the amount of burned
mass as a function of the density at which the burning took place.
If such graphs were similar for our simulations and the
phenomenological models, this would argue for the total chemical composition
of both models also being fairly equal, which in turn would be a strong hint
that light curves and spectra of both simulations might not be too different.
In other words, one would check whether characteristic properties of the
multidimensional numerical experiments agree with the results of 
the phenomenological models. If so, the predictions made by the 
phenomenological simulations could be believed to be valid also for 
our calculations.

Care must be taken that only quantities are chosen for comparison which
have a physical meaning in both models. For example, the time dependence of
the burning speed, which is often used to characterize phenomenological
models and is documented in many publications, cannot be used for this purpose,
since there exists no single burning speed in multidimensional simulations at
any given time. In this concrete situation, the time-dependent energy
generation rate would be the proper choice for a comparison, since 
it is well defined in both one- and multidimensional scenarios. It 
would therefore be convenient if the analysis of phenomenological
models contained global quantities like energy generation rates or
density dependence of burned mass, which can be easily compared to
multidimensional simulations.

\cite{khokhlov-00} describes
an alternative approach to model SN Ia explosions and presents
results of three-dimensional calculations. The numerical models employed
in his code differ quite strongly in various aspects from the schemes in our
code. In particular, the simulation is carried out on an adaptively refined
grid, the flame is modeled by a reaction-diffusion method, and its
propagation velocity is determined by the asymptotic rise speed of
Rayleigh-Taylor bubbles instead of the turbulent velocity fluctuations.

The initial setup described in his paper is centrally ignited, and no
perturbation is applied to the spherical flame surface. As soon as nonlinear
instabilities have developed, however, the explosion progresses
in a way which looks very similar to the 3D simulations
discussed in Sect.\ \ref{multipoint}.
This similarity applies to the energy production rate as well as the
geometrical features developed by the flame.

It is an encouraging fact that two independent and quite different attempts to
simulate a SN Ia without introducing artificial burning velocities and
chemical mixing yield fairly comparable results. This shows that the
explosion mechanism analysed here in detail, namely a pure
deflagration front in a C+O Chandrasekhar-mass white dwarf, is indeed
robust. 

\section{Discussion and outlook}
\label{discuss}

In this paper we have presented the results of three-dimensional numerical
simulations of thermonuclear deflagration fronts in Chandrasekhar mass white 
dwarfs composed of equal amounts of carbon and oxygen. We could show 
that independent of the details of the ignition process (which is
still far from being well understood) the white dwarf is always disrupted
by the release of nuclear energy. As far as we could check at present,
the properties of our models are in good agreement with observations
of typical type Ia supernovae. In particular the explosion energy
and the average chemical composition of the ejecta seem to fit the
observations (see also Table 1). This success of the models was
obtained without introducing any non-physical parameters, but just
on the basis of a physical and numerical model of subsonic turbulent 
burning fronts. We also stress that our models give clear evidence 
that the often postulated deflagration-detonation transition is not
needed to produce sufficiently powerful explosions.

\begin{table}[htbp]
  \centerline{
    \begin{tabular}{|l|c|c|c|}
    \hline
    model name & $m_{\text{Mg}}$ [$M_\odot$]&$m_{\text{Ni}}$ [$M_\odot$]&$E_{\text{nuc}}$ [$10^{50}$\,erg\vphantom{\raisebox{2pt}{\large A}}] \\
    \hline
    c3\_3d\_256 & 0.177 & 0.526 & 9.76 \\ \hline
    b5\_3d\_256 & 0.180 & 0.506 & 9.47 \\ \hline
    b9\_3d\_512 & 0.190 & 0.616 & 11.26\phantom{0} \\ \hline
    \end{tabular}
  }
  \caption{Overview over element production and energy release of all
    discussed supernova simulations}
  \label{burntable}
\end{table}

There are certainly several desirable additions and improvements.
The most crucial question still seems to be the ignition process,
although in 3D the dependence of the final outcome is
weaker than in our previous 2D models. In principle one could
try to simulate the ignition phase with
the numerical models we have developed, but because of the much
longer time scales this requires huge amounts of computer
time. First attempts in this direction are presently under way
(cf. \citealt{hoeflich-stein-2002}).

An improvement of the combustion model already
mentioned in \cite{reinecke-etal-99a}, i.e. the full reconstruction
of all thermodynamic quantities from their mean values in every
grid cell cut by the burning front, has now been completed and
was applied to the Landau-Darrieus instability \citep{roepke-etal-02}.
This new model should be implemented into the full code which,
in principle, seems to be possible but difficult. In any case,
this would increase the predictive power and reliability of the 
models.

In a similar direction, it might be worthwhile to test other
subgrid models in combination with our level set method. We do not
expect major changes because if another subgrid-model would give
us higher (or lower) burning speed on the grid scale this would
lead to more (or less) damping of small-scale structures leaving
the product of the two (and therefore the fuel consumption rate)
approximately unchanged. However, again, such studies are presently
in progress.

To conclude, we think that we have made another step towards
the understanding of type Ia supernovae.

\begin{acknowledgements}
We thank Sergei Blinnikov for pointing out the necessity to distinguish
between $^{56}$Ni and the remaining Fe-group elements, if our results
are to be used in light curve and spectra calculations.

This work was supported in part by the Deutsche Forschungsgemeinschaft under
Grant Hi 534/3-3.
The numerical computations were carried out on a Hitachi SR-8000 at the
Leibniz-Rechenzentrum M\"unchen as a part of the project H007Z.
\end{acknowledgements}

\bibliographystyle{apj}
\bibliography{refs}

\end{document}